\documentclass[journal,twocolumn]{IEEEtran}
\IEEEoverridecommandlockouts
\usepackage{cite}
\usepackage{amsmath,amssymb,amsfonts}
\usepackage{algorithmic}
\usepackage{graphicx}
\usepackage{textcomp}
\usepackage{xcolor}
\def\BibTeX{{\rm B\kern-.05em{\sc i\kern-.025em b}\kern-.08em
    T\kern-.1667em\lower.7ex\hbox{E}\kern-.125emX}}
   
   \usepackage[version=3]{mhchem} 
\usepackage{chemarr}

     \makeatletter
\newcommand\xleftrightarrow[2][]{%
  \ext@arrow 9999{\longleftrightarrowfill@}{#1}{#2}}
\newcommand\longleftrightarrowfill@{%
  \arrowfill@\leftarrow\relbar\rightarrow}
\makeatother

\begin{document}

\title{Time-Domain and Frequency-Domain Mappings of Voltage-to-Charge and Charge-to-Voltage  in Capacitive Devices}

\author{
\IEEEauthorblockN{Anis Allagui$^{1,2}$, Ahmed S. Elwakil$^{3,4}$, \emph{Senior Member IEEE} and Chunlei Wang$^2$, }\\
\IEEEauthorblockA{
{$^1$Dept. of Sustainable and Renewable Energy Engineering, University of Sharjah, United Arab Emirates} \\ 
{$^2$Dept. of Mechanical and Materials Engineering, Florida International University, Miami, FL33174, USA}\\
{$^3$Dept. of Electrical Engineering, University of Sharjah,  United Arab Emirates}\\
{$^4$Dept. of Electrical and Software Engineering, University of Calgary, Alberta, Canada (elwakil@ieee.org)}}
}
 
 \maketitle

\linespread{1.25}
 
\begin{abstract}

In this work, we aim to show that there are generally four possible mapping functions that can be used to map the time-domain or frequency-domain  representations of  an applied voltage input  to the resulting time-domain or frequency-domain electrical charge output; i.e. when the capacitive device is voltage-charged. Alternatively, there are four more possible combinations when the device is current-charged. The dual relationship between each pair of functions for the case of voltage or charge input are provided in terms of single or double Fourier transforms. All eight system functions coincide with each other if and only if a constant time- and frequency-independent capacitance is considered.

 

\end{abstract}

\vspace{5pt}


\section{Introduction}

Ideal electrostatic capacitors are constructed from two conductors separated by a perfect insulator. 
 The standard charge-voltage relation $q=cv$ represents the linear relationship between  the magnitude of the charge on each conductor  with the difference of potential  applied across the two conductors. 
 It is true in both time  and frequency domains for ideal capacitors having constant, time-independent and frequency-independent capacitances that can be only computed, and not measured, from   voltage and charge. 
 
 However, other capacitive devices, such as electrochemical capacitors (including electric double-layer capacitors (EDLCs) and pseudocapacitors), cannot be characterized with a constant capacitance value  \cite{westerlund1994capacitor, kumar2019analytical, shah2021analysis}. Their  electrodes are fractal and porous in nature, and are separated by ionic conductors leading to complex charge transports that are inherently different from ideal capacitors. 
 Coupled ionic and electronic transports  take place in these devices, which are ultimately controlled by diffusion dynamics and migration in the bulk electrolyte,  and through their micro/nano-porous  structured electrodes.  The correlated disorder of the background electrolyte also has an impact on the overall charge transport process which  makes it   deviate from normal diffusion to subdiffusion, and hence  internal dissipative processes \cite{morgado_relation_2002, allagui2021gouy}.  
 From a system-level point of view, 
   the spectral impedance phase angle of most -if not all- EDLCs and other similar energy storage devices deviate from the -90$^{\circ}$ value expected for ideal capacitors, even at near-dc, sub-Hz frequencies \cite{eis, shah2021analysis}. 
It has also been demonstrated experimentally that an EDLC charged with two different time-domain voltage waveforms to the same thermodynamic state of voltage and charge, $(v,q)$, exhibits different relaxation profiles when discharged into the same resistor \cite{memoryAPL}.  This means that there are inherent  memory effects  in these devices \cite{memQ, memoryAPL, allagui2021possibility}, and that the dynamic (dis)charging paths should be taken into consideration when evaluating their performance metrics   \cite{allagui2021inverse,fracorderreview, EC2015}. This  is not the case with ideal capacitors whose relaxation depends only on the initial condition. In addition, several time-domain transient measurements conducted on EDLCs show in some form or another  nonlinearities  and  power-law  behaviors that prohibit the extraction of a constant capacitance from the system's response \cite{EC2015,fracorderreview,  ragoisha2016false, QV}.  
    We note that the EDL structure for instance is ubiquitous in  many  other   fields of fundamental and applied electrochemistry such as in intercalation materials,   capacitive deionization and electrosorption,  colloidal suspensions, bipolar membranes, etc., and therefore properly evaluating its electrical parameters is paramount for meaningful progress and development in these research areas.


Unfortunately, there is a conceptual misunderstanding, even in specialized literature, on the difference in computation of performance metrics between an ideal capacitor and other capacitive energy storage devices 
 \cite{ieeeted}. 
 A classical example  is the direct calculation of capacitance from the time-domain expression $q=cv$, but the same formulation of charge equal to the product of capacitance by voltage   expressed in the frequency domain, and indirectly derived from impedance spectroscopy measurements, is used to characterize the same device, as if it was an ideal capacitor with constant capacitance (see \cite{EES, faraji2015development, devillers2014review, dupont2014large, ren2014kilohertz, muzaffar2019review, noori2019towards,  gharbi2020revisiting, taberna2003electrochemical, lockett2008differential, aoki2017decrease, wallar2017high} to cite a few). 
This is in direct contradiction with the convolution theorem \cite{doi:10.1021/acs.jpcc.1c01288}, and clearly indicates that there is an urgent need for a unified methodology of system identification for such   devices. 
 Most recently, in \cite{ieeeted} the authors studied the difference between the time-domain definition $q_t (t) =c_{t} (t)  v_t (t) $ and its frequency-domain counterpart defined as a convolution integral, i.e. $Q_t(f)=  ( V_{t} \circledast C_t ) \,(f)$, and the frequency-domain definition $ Q_f(f)=C_{f}(f)  \, V_f(f)$ and its time-domain equivalent $ q_f (t) = (c_{f}  \circledast  v_f ) \,(t)  $. 
 We denote by $V(f)$ the Fourier transform  of the input time-domain voltage $v(t)$ and by $Q(f)$ the Fourier transform of the output time-domain charge $q(t)$   
with the definitions $\mathcal{F}\{g(t)\} = G(f) = \int_{-\infty}^{\infty} g(t) e^{- j 2 \pi  f t} dt$ and $\mathcal{F}^{-1}\{G(t)\} = g(t) = (2\pi)^{-1} \int_{-\infty}^{\infty} G(f) e^{j 2 \pi f  t} dt$. In \cite{ieeeted}, we showed  that the extraction of capacitance functions (rather than constant capacitance) from these two pairs of equations for the case of dispersive mono-fractional-order  capacitors (known also as constant phase elements) subjected to  sinusoidal voltage excitation do not lead to  the same results. 
The same has been illustrated and verified for the case of linear voltage ramp and constant current input applied to a commercial supercapacitor device \cite{allagui2021inverse}. It is therefore one pair of equations or the other that can be used simultaneously, and the  relation $q=cv$ can only be applied on one-domain (time or frequency) but not the other.
We recommended nonetheless  the use of the multiplicative capacitance function in the frequency domain, i.e. $ Q_f(f)=C_{f}(f)  \, V_f(f)$,  in order to be in line with the definition of impedance of linear time-invariant systems \cite{ieeeted,doi:10.1021/acs.jpcc.1c01288}. 

This work is a direct extension of our previous studies \cite{ieeeted, allagui2021inverse}, and is primarily concerned with the system identification problem of capacitive devices from the two measurable quantities voltage and charge. The goal is to establish the different relationships between possible system capacitance functions \emph{without} the prior assumption of any model.  In Section II, we present the theoretical framework of this work while its implications are discussed further in Section III. It is important to note that there are no simulations or experimental results presented here since they have been readily presented in  \cite{ieeeted, allagui2021inverse} upon dealing with specific devices.

\section{Theory}

\begin{figure}[t]
\begin{center}
\includegraphics[width=.45\textwidth]{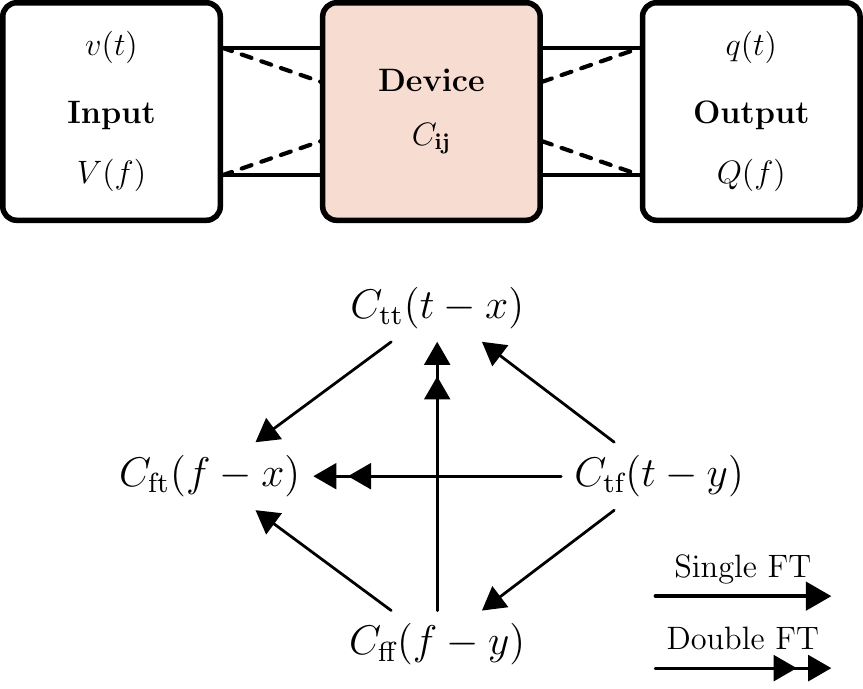}
\caption{Schematic diagram illustrating the two time-domain and frequency-domain representations of voltage input and the two time-domain and frequency-domain representations of the  charge output in a capacitive energy device. The different relationships between the capacitance functions $C_{\text{ij}}$ defined in Eq.\;\ref{eq4} are also summarized (FT: Fourier transform).}
\label{fig1}
\end{center}
\end{figure}

Consider the schematic illustration presented in Fig.\;\ref{fig1} relating the input/output of a capacitive device in the time and frequency domains. It is evident that there are four possible operators that can  map an input  voltage (in the time or frequency domain) to an output charge  (in the time or frequency-domain). When we refer to the frequency-domain, the time-domain quantity is dealt with indirectly through its Fourier frequency spectrum.    
 We can  write from the four signal representations $v(t)$, $V(f)$   and  $q(t)$, $Q(f)$ the following relations \cite{bello1964time, bello1963characterization}: 
\begin{subequations}
\begin{align}
 q(t) &= \mathcal{C}_{\text{tt}} \{v(t) \}\label{3a}  \\
 Q(f)  &= \mathcal{C}_{\text{ff}} \{ V(f)\} \\
 q(t)  &= \mathcal{C}_{\text{tf}} \{V(f)\} \label{3c} \\
 Q(f)  &= \mathcal{C}_{\text{ft}}\{ v(t)\}
\end{align}
\label{one}
\end{subequations}
where the four operators $\mathcal{C}_{\text{ij}}\;(\{i,j\}=\{\text{t},\text{f}\})$ map or transform the input signal in space $j=\{\text{t},\text{f}\}$ for time or frequency to space $i=\{\text{t},\text{f}\}$. For example, the operator $\mathcal{C}_{\text{tt}}$ in Eq.\;(\ref{3a}) represents the mapping relationship between the voltage input $v(t)$ (cause) and charge output $q(t)$ (effect) both defined in the time domain, whereas  the operator $\mathcal{C}_{\text{tf}}$ in Eq.\;(\ref{3c}) represents  the mapping relationship between the voltage input $V(f)$ defined in the frequency domain  and the charge output $q(t)$ defined in the time domain.  The mapping operators $\mathcal{C}_{\text{ij}}$ are unique, and can be in general related to   differential or integro-differential equations that govern the space-time charge distributions in a capacitive device. 
  
 For the case of devices which are assumed to be linear time-invariant   wherein  the input-output relationship is invariant under translation in time and the superposition principle holds, we can write the following linear integral equations of convolution \cite{rickard2005canonical}:
\begin{subequations}
\begin{align}
 q(t) &= \int C_{\text{tt}}(t-x) v(x) dx \label{4a} \\
 Q(f)  &= \int C_{\text{ff}}(f-y) V(y) dy \label{4b}\\
 q(t)  &= \int C_{\text{tf}}(t-y) V(y) dy \label{4c} \\
 Q(f)  &= \int C_{\text{ft}}(f-x) v(x) dx \label{4d}
\end{align}
\label{eq4}
\end{subequations}
The four kernels $C_{\text{ij}}\;(\{i,j\}=\{\text{t},\text{f}\})$ can be viewed as kernel  system functions  as proposed by Bello \cite{bello1964time, bello1963characterization}. 
Note that  when the limits of integration are not explicitly indicated in the equations they are to be taken from $-\infty$ to $\infty$, and the integrals are assumed to exist. Also, all quantities are treated as normalized  and thus dimensionless. 
  It is important to clarify at this point that in \cite{ieeeted} the focus  was  to establish the difference between assuming the correctness of the multiplicative equation $q=cv$  in the time domain versus assuming its correctness in the frequency domain, which are both widely used definitions in the literature. This led us to define four  different capacitance functions $c_t(t)$, $C_t(f)$ and $C_f(f)$, $c_f(t)$ by virtue of the convolution theorem as explained above.
In this work, we approach the problem from a general system identification perspective as described by the set of Eqs.\;(\ref{eq4}). Therefore the functions $C_{\text{ij}}\;(\{i,j\}=\{\text{t},\text{f}\})$ should not be confused with $c_t(t)$, $C_t(f)$, $C_f(f)$ or $c_f(t)$, which are not in conflict with the present analysis. 

Equations\;(\ref{4a}) and\;(\ref{4c}) provide two different expressions for the same time-domain charge $q(t)$. With the use of Parseval's equation   \cite{zemanian1987distribution} we can write:
\begin{equation}
\int_{-\infty}^{\infty} f(t) G(t) dt = \int_{-\infty}^{\infty} F(\omega) g(\omega) d\omega
\end{equation}
  (under the assumption that $f(t)$ and $g(t)$ are   absolutely integrable over $-\infty<t<\infty$). The kernel functions $C_{\text{tf}}(t-y)$ and $C_{\text{tt}}(t-x)$ constitute a Fourier transform pair with $t$ considered as a parameter:
\begin{equation}
\mathcal{F}\{ C_{\text{tf}}(t-y) \}=   C_{\text{tt}}(t-x)
\end{equation}
In the same way, we can establish from equating the frequency-domain charge $Q(f)$ given by both Eqs.\;(\ref{4b}) and \;(\ref{4d}) that: 
\begin{equation}
\mathcal{F}\{ C_{\text{ff}}(f-y) \}=   C_{\text{ft}}(f-x)
\end{equation} 
with $f$ being  considered as a parameter. 
From Eqs.\;(\ref{4a}), (\ref{4d}) 
(after applying the Fourier transform to both sides of Eq.\;(\ref{4a}))
and Eqs.\;(\ref{4c}), (\ref{4b}) (after applying the Fourier transform to both sides of Eq.\;(\ref{4c})), we can write:
\begin{equation}
C_{\text{ft}}(f-x) = \int e^{- j 2\pi f t} C_{\text{tt}}(t-x) dt
\end{equation}
and
\begin{equation}
C_{\text{ff}}(f-y) = \int e^{-j 2\pi f t} C_{\text{tf}}(t-y) dt
\end{equation}
respectively.
Finally,  
 from Eqs.\;(\ref{4a}) and \;(\ref{4b})  
 we can establish the following relation between $C_{\text{ff}}(f-y)$ and $C_{\text{tt}}(t-x)$:
\begin{align}
C_{\text{ff}}(f-y) &=\frac{1}{2\pi} \iint  C_{\text{tt}}(t-x) e^{-j 2\pi  t f} e^{j 2\pi y x} dt\, dx \nonumber \\
&=\frac{1}{2\pi} \iint  C_{\text{tt}}(t-x) e^{j 2\pi (yx - t f)} dt\, dx
\end{align}
and from Eqs.\;(\ref{4c}),\;(\ref{4d})  we have:
\begin{align}
C_{\text{ft}}(f-x) &=  \iint  C_{\text{tf}}(t-y) e^{-j 2\pi  t f} e^{-j 2\pi x y} dt\, dy \nonumber \\
&=  \iint  C_{\text{tf}}(t-y) e^{-j 2\pi ( t f  +xy) }  dt\, dy 
\end{align}

Figure\;\ref{fig1} summarizes the four different forms of system   functions $C_{\text{ij}}\;(\{i,j\}=\{\text{t},\text{f}\})$  defined in Eq.\;(\ref{eq4}) along with  their pair  relationships either as single or double Fourier transforms.

A simple example that demonstrates the correspondence between these system functions is the case in which the charge is  a delayed copy input voltage $v(t)$, i.e.   \cite{rickard2005canonical}:
\begin{equation}
 q(t)  = \mathcal{C}_{\text{tt}} \{ v(t) \} = A v(t-\tau_0)\quad(A=1)
\end{equation} 
 Thus:
 \begin{equation}
C_{\text{tt}}(t-x) = \delta (t-x-\tau_0)
\end{equation}
 which verifies Eq.\;\ref{4a}. From here the rest of the system functions in Eqs.\;\ref{4b}, \ref{4c} and \ref{4d} are 
 \begin{align}
   C_{\text{ff}}(f-y)& = \delta(f-y) e^{-j  2\pi  y \tau_0} \\
  C_{\text{tf}}(t-y) &= e^{j 2\pi  y (t-\tau_0)} \\
   C_{\text{ft}}(f-x)& = e^{-j 2\pi f(x+\tau_0)}
\end{align}
respectively. These functions can be transformed into one another   as indicated in Fig.\;\ref{fig1}. 
Another example is when the resulting time-domain charge $q(t)$ is a response to an input voltage $v(t)$   given by \cite{rickard2005canonical}:
\begin{equation}
 q(t)  = \mathcal{C}_{\text{tt}} \{ v(t) \} = v(t-\tau_0) e^{j 2 \pi  f_0 t}
\end{equation} 
The four capacitance functions are then:
 \begin{align}
 C_{\text{tt}}(t-x) & = \delta (t-x-\tau_0) e^{j 2 \pi  f_0 t}\\
   C_{\text{ff}}(f-y)& = \delta(f-y -f_0) e^{- j 2\pi  y \tau_0} \\
  C_{\text{tf}}(t-y) &= e^{2\pi j  t (y+f_0)} e^{-j 2 \pi y \tau_0 } \\
   C_{\text{ft}}(f-x)& = e^{-j 2\pi (f-f_0)(x+\tau_0)}
\end{align}
 Such types of functions are not only necessary in relating charge and voltage in capacitive devices but are also necessary in modeling communication channels for instance. As an example, the intrabody communication channel is predominantly capacitive and modeled by distributed networks of ideal capacitors, as shown in \cite{xu2019modeling} for an in-vehicle scenario. The availability of frequency domain-based network analyzers forces the application of frequency-domain modeling techniques as seen for example in the scattering matrix results reported in \cite{petrillo2017human}.

\begin{figure}[b]
\begin{center}
\includegraphics[width=.45\textwidth]{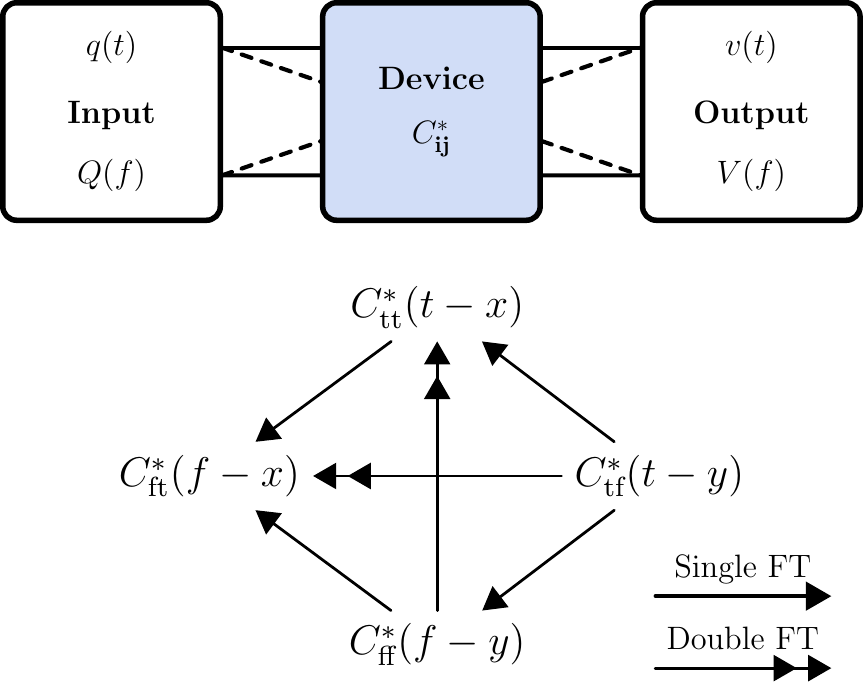}
\caption{Schematic diagram similar to Fig.\;\ref{fig1} with the electrical charge being the input and the voltage being the output in a capacitive energy device.}
\label{fig2}
\end{center}
\end{figure}

The equation pairs presented above are all related to the case when a capacitive device is being voltage-charged. The electric charge on the device is thus a result (output). However, when using the electrical current instead of voltage to charge capacitive devices, this implies that the charge is readily created by the charging power supply since $dq = \int i(t)dt$ \cite{2018-1} before being applied on the device. The input  in this case is the electrical charge and the voltage measured is thus a result (output), as shown in Fig.\;\ref{fig2}. A new set of four equations similar to Eqs.\;(\ref{one}) can be deduced for this case albeit the LHS must be the voltage and the RHS must be the charge, with  the mapping functions ($C^*_{\text{ij}}$) being inverse-capacitance functions. 
 When a device is reciprocal, the output charge from Fig.\;\ref{fig1} fed as an input in Fig.\;\ref{fig2} should result in an output voltage from Fig.\;\ref{fig2} identical to that supplied as input to Fig.\;\ref{fig1} \cite{2018-1}. However, this is not always necessarily true and hence the computation of a \emph{capacitance} function is associated only with voltage excitation whereas the computation of \emph{inverse capacitance} functions should be done in the case of current excitations.

\section{Discussion}

A capacitive device is treated here as a lumped electrical signal processing element  at the link between voltage input and charge output or charge input and voltage output (single-input single-output device) as depicted in Figs.\;\ref{fig1} and\;\ref{fig2}, respectively. The charge, being the time-integral of the current, can be the result of all current contributions taking place within the device and thus there is no real discrimination between what is effectively being used for the purpose of energy storage for future reuse, or what is being dissipated. In mathematical terms, the whole device can be thought of as an operator which transforms a certain input signal into an output signal, both of  which can be described either in the time domain or frequency domain \cite{bello1964time}. 
Cyclic voltammetry, constant current charge/discharge and impedance spectroscopy are all  
routine  techniques  for time-domain and frequency-domain measurements  of capacitive  energy storage devices \cite{allagui2021inverse,ieeeted,EC2015, conv}.  
However, the relations presented above are actually valid for any arbitrary excitations as long as the linearity and time-invariance of the device are respected, noting   that there are also   identification methods for  time-variant and nonlinear systems \cite{bello1963characterization, szekeres2021methods, hallemans2021best}.   
Furthermore, in reality the input and output signals can have superposed noises associated with measurement errors or other environmental conditions that should be taken into account when estimating  the corresponding mapping system functions. This requires the use of efficient regularization methods and inverse problem computations and not simple multiplications and divisions as commonly done using the ideal capacitor formul\ae.  

In fact, from the classical treatment inherited from ideal capacitors  one can write interchangeably $q=cv$ or $v=q/c$  without worrying about which is the input variable (voltage or charge). This is clearly incorrect for the general case of  capacitive  functions.  There is indeed a  difference between  the system functions $C_{\text{ij}}$ and $C^*_{\text{ij}}$  for a linear  time-invariant capacitive device, and thus between  voltage charging and current charging as illustrated in Figs.\;\ref{fig1} and \;\ref{fig2}.   
The results  further highlight that  computing the  capacitive (or inverse capacitive) functions   is dependent on the form and type of the applied excitation, which is different from ideal capacitors. Applying a step function, linear ramp or sinusoidal excitation (voltage or current)  will result in different capacitive functions. Therefore, one should  specify clearly how the device is being excited and which capacitive  (or inverse capacitive) function  is being computed    because these functions cannot be identical to each other  except for the case of ideal capacitors (i.e. $C_{\text{ij}} = C^*_{\text{ij}} = c$). We believe that this will have significant implications on the way capacitance (and similarly inductance) is computed in the future in an era where capacitive energy storage devices are on high demand in electrical vehicles, IoT systems and many other applications.
 
\section{Conclusion}

This work is an important companion to our recent work in  \cite{ieeeted}. We have shown here  that there are several  system functions that can map an arbitrary input voltage/charge, represented in time or frequency, to an output charge/voltage, represented in time or frequency for a capacitive device (see Figs.\;\ref{fig1} and \;\ref{fig2}). 
 From the estimation of one function (for example $ {C}_{\text{tt}}$ that maps voltage to charge when they are both defined in the time domain, one can deduct all other functions  using single or double  Fourier transforms. 
 We emphasize that the functions $C_{\text{ij}}$ and $C^*_{\text{ij}}$ should be regarded as system functions and not capacitances, which are solely characteristics of ideal capacitors.

\section*{Acknowledgements}
\small {This work is supported by NSF project \#2126190 (C.W \& A.A.)}


\bibliographystyle{IEEEtran}


\begin{thebibliography}{10}
\providecommand{\url}[1]{#1}
\csname url@samestyle\endcsname
\providecommand{\newblock}{\relax}
\providecommand{\bibinfo}[2]{#2}
\providecommand{\BIBentrySTDinterwordspacing}{\spaceskip=0pt\relax}
\providecommand{\BIBentryALTinterwordstretchfactor}{4}
\providecommand{\BIBentryALTinterwordspacing}{\spaceskip=\fontdimen2\font plus
\BIBentryALTinterwordstretchfactor\fontdimen3\font minus
  \fontdimen4\font\relax}
\providecommand{\BIBforeignlanguage}[2]{{%
\expandafter\ifx\csname l@#1\endcsname\relax
\typeout{** WARNING: IEEEtran.bst: No hyphenation pattern has been}%
\typeout{** loaded for the language `#1'. Using the pattern for}%
\typeout{** the default language instead.}%
\else
\language=\csname l@#1\endcsname
\fi
#2}}
\providecommand{\BIBdecl}{\relax}
\BIBdecl

\bibitem{westerlund1994capacitor}
S.~Westerlund and L.~Ekstam, ``Capacitor theory,'' \emph{IEEE Trans. Dielectr.
  Electr. Insul.}, vol.~1, no.~5, pp. 826--839, 1994.

\bibitem{kumar2019analytical}
M.~R. Kumar, S.~Ghosh, and S.~Das, ``Analytical formulation for power, energy,
  and efficiency measurement of ultracapacitor using fractional calculus,''
  \emph{IEEE Trans. Instrum. Meas.}, vol.~68, no.~12, pp. 4834--4844, 2019.

\bibitem{shah2021analysis}
Z.~M. Shah and F.~A. Khanday, ``Analysis of disordered dynamics in polymer
  nanocomposite dielectrics for the realization of fractional-order
  capacitor,'' \emph{IEEE Trans. Dielectr. Electr. Insul.}, vol.~28, no.~1, pp.
  266--273, 2021.

\bibitem{morgado_relation_2002}
R.~Morgado, F.~A. Oliveira, G.~G. Batrouni, and A.~Hansen, ``Relation between
  {Anomalous} and {Normal} {Diffusion} in {Systems} with {Memory},''
  \emph{Phys. Rev. Lett.}, vol.~89, no.~10, 2002.

\bibitem{allagui2021gouy}
A.~Allagui, H.~Benaoum, and O.~Olendski, ``On the gouy-chapman-stern model of
  the electrical double-layer structure with a generalized boltzmann factor,''
  \emph{Physica A}, p. 126252, 2021.

\bibitem{eis}
A.~Allagui, A.~S. Elwakil, B.~J. Maundy, and T.~J. Freeborn, ``Spectral
  capacitance of series and parallel combinations of supercapacitors,''
  \emph{ChemElectroChem}, vol.~3, no.~9, pp. 1429--1436, 2016.

\bibitem{memoryAPL}
A.~Allagui, D.~Zhang, and A.~S. Elwakil, ``Short-term memory in electric
  double-layer capacitors,'' \emph{Appl. Phys. Lett.}, vol. 113, pp.
  253\,901--5, 2018.

\bibitem{memQ}
A.~Allagui, D.~Zhang, I.~Khakpour, A.~S. Elwakil, and C.~Wang, ``Quantification
  of memory in fractional-order capacitors,'' \emph{J. Phys. D}, vol.~53, no.
  02LT03, 2020.

\bibitem{allagui2021possibility}
A.~Allagui and A.~S. Elwakil, ``Possibility of information encoding/decoding
  using the memory effect in fractional-order capacitive devices,'' \emph{Sci.
  Rep.}, vol.~11, no.~1, pp. 1--7, 2021.

\bibitem{allagui2021inverse}
A.~Allagui and M.~E. Fouda, ``Inverse problem of reconstructing the capacitance
  of electric double-layer capacitors,'' \emph{Electrochim. Acta}, p. 138848,
  2021.

\bibitem{fracorderreview}
A.~Allagui, T.~J. Freeborn, A.~S. Elwakil, M.~E. Fouda, B.~J. Maundy, A.~G.
  Radwanh, Z.~Said, and M.~A. Abdelkareem, ``Review of fractional-order
  electrical characterization of supercapacitors,'' \emph{J. Power Sources},
  vol. 400, no. 457--467, 2018.

\bibitem{EC2015}
A.~Allagui, T.~J. Freeborn, A.~S. Elwakil, and B.~J. Maundy, ``Reevaluation of
  performance of electric double-layer capacitors from constant-current
  charge/discharge and cyclic voltammetry,'' \emph{Sci. Rep.}, vol.~6, no.
  38568, 2016.

\bibitem{ragoisha2016false}
G.~Ragoisha and Y.~Aniskevich, ``False capacitance of supercapacitors,''
  \emph{arXiv preprint arXiv:1604.08154}, 2016.

\bibitem{QV}
M.~E. Fouda, A.~Allagui, A.~S. Elwakil, S.~Das, C.~Psychalinos, and A.~G.
  Radwan, ``Nonlinear charge-voltage relationship in constant phase element,''
  \emph{AEU Int. J. Electron. Commun.}, vol. 117, no. 1533104, 2020.

\bibitem{ieeeted}
A.~Allagui, A.~S. Elwakil, and M.~E. Fouda, ``Revisiting the time-domain and
  frequency-domain definitions of capacitance,'' \emph{IEEE Trans. Electron
  Devices}, vol.~68, no.~6, 2021.

\bibitem{EES}
M.~Zhang, X.~Yu, H.~Ma, W.~Du, L.~Qu, C.~Li, and G.~Shi, ``Robust graphene
  composite films for multifunctional electrochemical capacitors with an
  ultrawide range of areal mass loading toward high-rate frequency response and
  ultrahigh specific capacitance,'' \emph{Energy Environ. Sci.}, vol.~11,
  no.~3, pp. 559--565, 2018.

\bibitem{faraji2015development}
S.~Faraji and F.~N. Ani, ``The development supercapacitor from activated carbon
  by electroless plating---a review,'' \emph{Renewable Sustainable Energy
  Rev.}, vol.~42, pp. 823--834, 2015.

\bibitem{devillers2014review}
N.~Devillers, S.~Jemei, M.-C. P{\'e}ra, D.~Bienaim{\'e}, and F.~Gustin,
  ``Review of characterization methods for supercapacitor modelling,'' \emph{J.
  Power Sources}, vol. 246, pp. 596--608, 2014.

\bibitem{dupont2014large}
M.~F. Dupont, A.~F. Hollenkamp, and S.~W. Donne, ``Large amplitude
  electrochemical impedance spectroscopy for characterizing the performance of
  electrochemical capacitors,'' \emph{J. Electrochem. Soc.}, vol. 161, no.~4,
  p. A648, 2014.

\bibitem{ren2014kilohertz}
G.~Ren, X.~Pan, S.~Bayne, and Z.~Fan, ``Kilohertz ultrafast electrochemical
  supercapacitors based on perpendicularly-oriented graphene grown inside of
  nickel foam,'' \emph{Carbon}, vol.~71, pp. 94--101, 2014.

\bibitem{muzaffar2019review}
A.~Muzaffar, M.~B. Ahamed, K.~Deshmukh, and J.~Thirumalai, ``A review on recent
  advances in hybrid supercapacitors: Design, fabrication and applications,''
  \emph{Renewable Sustainable Energy Rev.}, vol. 101, pp. 123--145, 2019.

\bibitem{noori2019towards}
A.~Noori, M.~F. El-Kady, M.~S. Rahmanifar, R.~B. Kaner, and M.~F. Mousavi,
  ``Towards establishing standard performance metrics for batteries,
  supercapacitors and beyond,'' \emph{Chem. Soc. Rev.}, vol.~48, no.~5, pp.
  1272--1341, 2019.

\bibitem{gharbi2020revisiting}
O.~Gharbi, M.~T. Tran, B.~Tribollet, M.~Turmine, and V.~Vivier, ``Revisiting
  cyclic voltammetry and electrochemical impedance spectroscopy analysis for
  capacitance measurements,'' \emph{Electrochim. Acta}, p. 136109, 2020.

\bibitem{taberna2003electrochemical}
P.~Taberna, P.~Simon, and J.-F. Fauvarque, ``Electrochemical characteristics
  and impedance spectroscopy studies of carbon-carbon supercapacitors,''
  \emph{J. Electrochem. Soc.}, vol. 150, no.~3, p. A292, 2003.

\bibitem{lockett2008differential}
V.~Lockett, R.~Sedev, J.~Ralston, M.~Horne, and T.~Rodopoulos, ``Differential
  capacitance of the electrical double layer in imidazolium-based ionic
  liquids: influence of potential, cation size, and temperature,'' \emph{J.
  Phys. Chem. C}, vol. 112, no.~19, pp. 7486--7495, 2008.

\bibitem{aoki2017decrease}
K.~J. Aoki, J.~Chen, X.~Zeng, and Z.~Wang, ``Decrease in the double layer
  capacitance by faradaic current,'' \emph{RSC advances}, vol.~7, no.~36, pp.
  22\,501--22\,509, 2017.

\bibitem{wallar2017high}
C.~Wallar, R.~Poon, and I.~Zhitomirsky, ``High areal capacitance of
  v2o3--carbon nanotube electrodes,'' \emph{J. Electrochem. Soc.}, vol. 164,
  no.~14, p. A3620, 2017.

\bibitem{doi:10.1021/acs.jpcc.1c01288}
A.~Allagui, A.~S. Elwakil, and H.~Eleuch, ``Highlighting a common confusion in
  the computation of capacitance of electrochemical energy storage devices,''
  \emph{J. Phys. Chem. C}, vol. 125, pp. 9591--9592, 2021.

\bibitem{bello1964time}
P.~Bello, ``Time-frequency duality,'' \emph{IEEE Trans. Inf. Theory}, vol.~10,
  no.~1, pp. 18--33, 1964.

\bibitem{bello1963characterization}
------, ``Characterization of randomly time-variant linear channels,''
  \emph{IEEE Trans. Commun. Syst.}, vol.~11, no.~4, pp. 360--393, 1963.

\bibitem{rickard2005canonical}
S.~T. Rickard, R.~V. Balan, H.~V. Poor, and S.~Verd{\'u}, ``Canonical
  time-frequency, time-scale, and frequency-scale representations of
  time-varying channels,'' \emph{Commun. Inf. Syst.}, vol.~5, pp. 197--226,
  2005.

\bibitem{zemanian1987distribution}
A.~H. Zemanian, \emph{Distribution theory and transform analysis: an
  introduction to generalized functions, with applications}.\hskip 1em plus
  0.5em minus 0.4em\relax Courier Corporation, 1987.

\bibitem{xu2019modeling}
Y.~Xu, Z.~Huang, S.~Yang, Z.~Wang, B.~Yang, and Y.~Li, ``Modeling and
  characterization of capacitive coupling intrabody communication in an
  in-vehicle scenario,'' \emph{Sensors}, vol.~19, no.~19, p. 4305, 2019.

\bibitem{petrillo2017human}
L.~Petrillo, J.~Sarrazin, H.~Libotte, A.~Benlarbi-Delai, F.~Horlin, and
  P.~De~Doncker, ``Human body communication channel modeling using vector
  network analyzer measurement,'' in \emph{2017 11th European Conference on
  Antennas and Propagation (EUCAP)}.\hskip 1em plus 0.5em minus 0.4em\relax
  IEEE, 2017, pp. 1868--1870.

\bibitem{2018-1}
A.~Allagui, A.~S. Elwakil, and T.~J. Freeborn, ``Supercapacitor reciprocity and
  response to linear current and voltage ramps,'' \emph{Electrochim. Acta},
  vol. 258, pp. 1081--1085, 2017.

\bibitem{conv}
M.~E. Fouda, A.~S. Elwakil, A.~Allagui, H.~Rezk, and A.~M. Nassef,
  ``Convolution-based estimation of supercapacitor parameters under periodic
  voltage excitations,'' \emph{J. Electrochem. Soc.}, vol. 166, no.~10, pp.
  A2267--A2269, 2019.

\bibitem{szekeres2021methods}
K.~J. Szekeres, S.~Vesztergom, M.~Ujv{\'a}ri, and G.~G. L{\'a}ng, ``Methods for
  the determination of valid impedance spectra in non-stationary
  electrochemical systems: Concepts and techniques of practical importance,''
  \emph{ChemElectroChem}, vol.~8, no.~7, pp. 1233--1250, 2021.

\bibitem{hallemans2021best}
N.~Hallemans, R.~Pintelon, E.~Van~Gheem, T.~Collet, R.~Claessens, B.~Wouters,
  K.~Ramharter, A.~Hubin, and J.~Lataire, ``Best linear time-varying
  approximation of a general class of nonlinear time-varying systems,''
  \emph{IEEE Trans. Instrum. Meas.}, 2021.

\end{thebibliography}


\end{document}